\documentclass[preprint, APS]{revtex4}

\usepackage{graphicx}
\usepackage{graphicx}
\usepackage{hyperref}
\usepackage{mathtools}
\usepackage{amsmath}
\usepackage{natbib}
\usepackage{color}

\begin{document}

\title{2D molecular magnets with weak topological invariant magnetic moments:~ Mathematical prediction of targets for chemical synthesis}

\author{Daniel M. Packwood$^1$}
\email{packwood@wpi-aimr.tohoku.ac.jp}

\author{Kelley T. Reaves$^{1,2}$}

\author{Filippo L. Federici$^{1,3}$}

\author{Helmut G. Katzgraber$^{2,4}$}

\author{Winfried Teizer$^{1,2,4}$}

\affiliation{$^1$WPI-Advanced Institute for Materials Research, Tohoku University, 2-1-1, Katahira, Aoba-ku, Sendai 980-8577, Japan\\
$^2$Materials Science and Engineering, Texas A\&M University, College Station, TX 77843-3003, USA\\
$^3$Department of Physics and Astronomy, University College London, Gower Street, London WC1E-6BT, United Kingdom\\
$^4$Department of Physics and Astronomy, Texas A\&M University, College Station, TX 77843-4242, USA\\}

\begin{abstract}
An open problem in applied mathematics is to predict interesting molecules which are realistic targets for chemical synthesis. In this paper, we use a spin Hamiltonian-type model to predict molecular magnets (MMs) with magnetic moments that are intrinsically robust under random shape deformations to the molecule. Using the concept of convergence in probability, we show that for MMs in which all spin centers lie in-plane and all spin center interactions are ferromagnetic, the total spin of the molecule is a `weak topological invariant' when the number of spin centers is sufficiently large. By weak topological invariant, we mean that the total spin of the molecule only depends upon the arrangement of spin centers in the molecule, and is unlikely to change under shape deformations to the molecule. Our calculations show that only between 20 and 50 spin centers are necessary for the total spin of these MMs to be a weak topological invariant. The robustness effect is particularly enhanced for 2D ferromagnetic MMs that possess a small number of spin rings in the structure. Our results therefore give reasonable targets for synthetic chemistry, and may help identify MMs that have intact magnetic properties upon deposition onto metal surfaces.
\end{abstract}

\maketitle

\section{Introduction}

Compared to physics or even biology, the application of non-elementary mathematics to problems in chemistry remains relatively unexplored. The purpose of this paper is to show how analytical concepts such as convergence in probability and rates of convergence can be used to predict interesting new molecules that are reasonable targets for chemical synthesis. As a convenient example, we will consider a class of molecules called molecular magnets (MMs). MMs are large molecules containing 4 to 100 electron spin centers (typically the d-electrons of transition metal ions) coupled in such a way that the molecule has a large total electron spin and an appreciable magnetic moment \cite{FriedmanSarachik, SessoliNature, WinneypennyNEW, GatteschiNEW}. MMs possess unique magnetic properties that span both classical and quantum domains, making them of great fundamental interest, and the elaborate structures of large MMs make them challenging targets for synthetic chemistry \cite{Winpenny}. Figure 1 shows the structure of the prototypic example of a MM, Mn$_{12}$(acetate) [formally Mn$_{12}$O$_{12}$(O$_{2}$CCH$_{3}$)$_{16}$(H$_2$O)$_4$]. Mn$_{12}$(acetate) consists of four Mn$^{4+}$ ions and eight Mn$^{3+}$ ions connected together through Mn-O-Mn or Mn-O-O-Mn bridges. The coupling between the electrons of the Mn ions \textit{via} the Mn-O-Mn and Mn-O-O-Mn bridges gives the Mn$_{12}$(acetate) molecule a total electron spin of 10 \cite{SessoliNature, Sta08}. The structure of MMs are often described in terms of the arrangement of spin centers in the molecule. In the case of Mn$_{12}$(acetate), the four Mn$^{4+}$ ions are arranged in a tetrahedral `core' with the eight Mn$^{3+}$ ions arranged in a ring around the periphery of the tetrahedron. Many other types of MMs with a wide range of spin center arrangements have also been characterised in the literature. A small sample includes the tetrahedral, cubic, and planar Mn$_4$ and lanthanide-4 complexes \cite{Yang2003, Liu2009, Nguyen2011, Liu2012}, the cage-like Fe$_{15}$ and Fe$_{36}$ complexes \cite{Ren2013, Beavers2013}, the Fe$_{10}$ ring \cite{Fe10}, the dual-tetrahedral Fe$_{7}$  complex \cite{DualTD}, and the large torus-shaped Mn$_{84}$ complex \cite{AromiBrechin2006, Tasiopoulos2004}. 

\begin{figure}[h]
	\centering
		\includegraphics[width=0.5\textwidth]{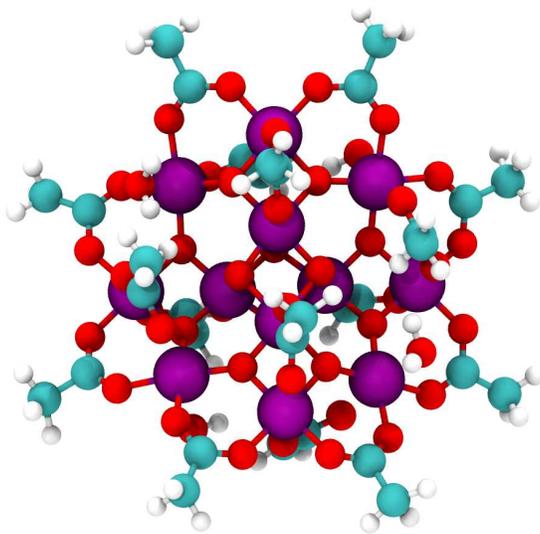}
	\caption{Structure of the Mn$_{12}$(acetate) molecule. The large, purple spheres are Mn ions, the red spheres are O atoms, the light blue spheres are C atoms, and the white spheres are H atoms. The radius of the Mn atoms has been enlarged from their usual ionic radius for clarity.}
	\label{fig:Figure1}
\end{figure}

A key problem that is currently facing the MM community is how to bind MMs to conducting surfaces or thin films in such a way that the magnetic properties of the molecule remain intact. Adsorption of MMs to surfaces is necessary for technological applications of MMs, and also to study their magnetization properties in scanning tunneling microscopy experiments \cite{Gatteschi2009, Reaves2013}. However, the equilibrium gas-phase shape of a MM is not expected to be maintained upon adsorption due to various uncontrollable factors, such as the orientation of the molecule upon collision with the surface, the small-scale roughness of the surface, and the thermal motions of the atoms of the surface. Experimental and quantum chemical studies have shown that the magnetic polarisation of Mn$_{12}$(acetate) is reduced when adsorbed to metallic surfaces due to oxidation of the Mn ions and structural deformations of the molecule \cite{Mannini2008, Park, Grumbach2009}. The magnetic properties of a different Mn$_{12}$ complex were also reported to be affected when the molecule was adsorbed on a gold surface with a self-assembled monolayer \cite{Moro}. Other studies have further shown that significant structural deformations and changes in the spin center-spin center coupling strengths occur when Cr$_7$Ni and Mn$_{6}$ MMs are adsorbed on gold surfaces \cite{CorradiniRings2, CorradiniRings, Totti2013}. On the other hand, some interesting experiments have achieved surface-adsorbed MMs with intact magnetic properties. For example, it has been demonstrated that by a careful choice of deposition method and substrate surface, the magnetic properties of Mn$_{12}$(acetate) could be preserved upon adsorption to a surface \cite{Kahle2012, Sun2013ex}. Moreover, it was recently reported that the magnetic properties of certain Fe$_4$ and Fe$_3$Cr complexes are relatively robust upon adsorption to a gold surface \cite{ManniniNM2009, ManniniNL2010, ManniniInorganic}. While these experimental advancements are promising, it is still desirable to identify other MMs with magnetic properties that are intrinsically stable to shape deformations to the molecule.

Let us consider the task of using a mathematical model to predict structures for MMs with robust magnetic properties under shape deformations. In order to achieve this, we need a model that is sufficiently idealised yet still realistic enough to capture key molecular properties of interest. MMs are a rare class of molecules for which such models - spin Hamiltonians - are available. In its simplest form, the spin Hamiltonian can be written as

\begin{equation}
\hat{H} = -\sum_{i,j} J_{ij} \hat{\mathbf{s}}_i \cdot \hat{\mathbf{s}}_j,
\end{equation}
  
\noindent
where the sum runs over all pairs of interacting spin centers (e.g., Mn ions) in the molecule, $J_{ij}$ measures the coupling strength (exchange energy) between spin centers $i$ and $j$, and $\hat{\mathbf{s}}_i$ is the spin operator for spin center $i$. Despite their simplicity, spin Hamiltonian models provide a qualitatively correct description of the magnetic properties of MMs \cite{EngelhardtSchroder, AmericanJofPhys} and with certain additional terms are even capable of describing some exotic magnetic phenomena such as quantum tunneling of the magnetization in the Mn$_{12}$(acetate) molecule \cite{GatteschiSessoli}. In spin Hamiltonian models, much of the detailed physics of the MM is condensed into the coupling constants $J_{ij}$. This prevents spin Hamiltonian models from describing certain phenomena such the origin of the spin center-spin center coupling in an MM (such as the superexchange mechanism through the oxygen atoms \cite{Yam58, Kon59}) or how the coupling strengths are affected by the molecular structure of the MM (such as the organic groups attached to the O atoms in Mn$_{12}$-type molecules \cite{Han04,Bou10}). Spin Hamiltonian models are therefore potentially useful if we are strictly interested in predicting MMs with interesting magnetic properties, such as robustness of the total spin under molecular shape deformations.

In this paper, we use a modification of the spin Hamiltonian model to show that for a certain class of MMs, the total spin behaves like a weak topological invariant when the number of electron spin centers in the molecule is sufficiently large and the molecule is in the ground state of the total spin. By `weak topological invariant', we mean a physical property that only depends upon the arrangement of spin centers in the equilibrium molecule and does not change when the equilibrium shape of the molecule undergoes a small, random deformation, provided that the deformations do not alter the chemical structure of the molecule. Such random deformations might arise from the thermal motion of the adsorbent surface or from adsorption onto a particularly rough region of the surface. The prefix `weak' indicates that there may exist deformations that do cause large changes to the total spin but have a low probability of occurring. While our definition of weak topological invariant is more restrictive than the usual definition of topological invariant, it is still a sufficient criterion for identifying robust molecules. The class of MMs that are considered are where all spin center-spin center interactions in the molecule are ferromagnetic, all distances between interacting spins centers are equal, and all spins centers lie in a two-dimensional plane (2D ferromagnetic MMs). While these are significant idealisations compared to realistic MMs, our model can be extended to include antiferromagnetic interactions and varying distances between interacting spin centers. However, it turns out that for 2D ferromagnetic MMs the weak topological invariant property appears quite naturally from the model. We will see that when the molecule contains a relatively small number of `spin rings' (i.e., interacting spin centers arranged in a ring shape), only between 20 and 50 spins are necessary for the total spin to behave like a weak topological invariant. While such large 2D MMs do not appear to have been reported in the chemistry literature, they are within the size range of most three dimensional MMs that have been reported (between about 4 and 100 spins). In particular, we find that ferromagnetic MMs composed of only one spin ring and around 20 spins have total spins that are particularly robust to shape deformations. Smaller rings composed of 8 - 10 metal ions are well-known in the MM literature \cite{Fe10, Cr8}, and therefore the `large' spin rings that we predict here may be reasonable targets for chemical synthesis. As will be seen, our definition of weak topological invariant is closely related to the mathematical concept of convergence in probability of a random variable (here representing the stability of the total spin) to zero, and the task of predicting 2D MMs with weak topological invariant total spins is related to the important problem of bounding rates of convergence in probability. This work might therefore be regarded as a kind of `mathematical chemistry', i.e., the prediction of interesting molecules using non-elementary mathematical concepts.

\section{Model}

We first develop the model for the general case where the spins of the MM are not restricted to a two dimensional plane. The equilibrium structure of the MM is represented as collection of points (`vertices') in a three dimensional Euclidean space $\mathbf{R}^3$, with each vertex representing the location of an electron spin center in the MM (e.g., Mn ions). Edges are drawn between vertices that correspond to interacting spin centers (e.g., Mn-O-Mn bridges). Only nearest-neighbor interactions are considered. Vertex $i$ lies at point $\mathbf{r}_i$ , and the length of the edge between vertex $i$ and vertex $j$ is $r_{ij}$. We assume constant edge lengths, i.e., that $r_{ij} = r = 1$ (in appropriate units) for all interacting spin centers. The strength of the interaction between spin center $i$ and spin center $j$ is described by associating a function $J_{ij} = J(r_{ij})$ with the edge $(i,j)$. $J_{ij}$ describes the exchange interaction between the electrons at spin centers $i$ and $j$. Because all edges have constant length, the functions $J_{ij}$ are independent of $i$ and $j$ and are equal to $J(r)$. We assume that all interactions between spin centers are ferromagnetic, i.e., that $J(r) > 0$. Shape deformations to the molecule are modeled by adding random vectors $\mathbf{X}_1, \mathbf{X}_2, \ldots$ to the corresponding position vectors $\mathbf{r}_1, \mathbf{r}_2, \ldots$. The components of $\mathbf{X}_1, \mathbf{X}_2, \ldots$ are independent Gaussian random variables with mean 0 and variance $\sigma^2_x$. Following the deformation, the interaction strength between spin centers $i$ and $j$ changes by an amount $\Delta J_{ij} = J(r+\Delta r_{ij}) - J(r)$, where $\Delta r_{ij}$ is the change in the distance between vertices $i$ and $j$ following the deformation. The deformations are assumed to be first-order in $\Delta r_{ij}$. A simplified sketch of the model is shown in Figure 2. A deformation roughly resembles the procedure used in Monte Carlo simulations of the compressible Ising model \cite{Cannavacciuolo}.

\begin{figure}
	\centering
		\includegraphics[width=0.75\textwidth]{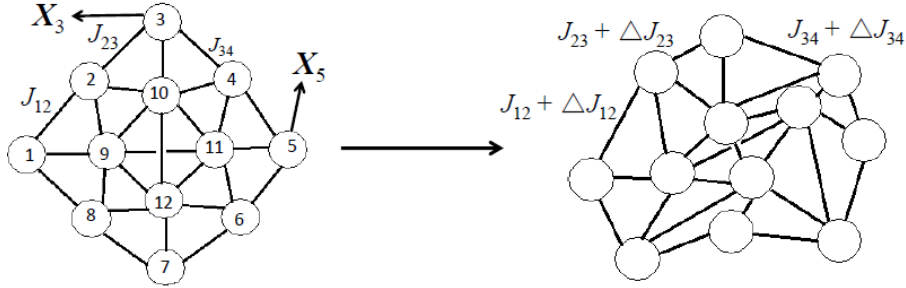}
	\caption{Sketch of the model for an Mn$_{12}$-like MM undergoing a random shape deformation. Only some of the interaction constants $J_{ij}$ and random vectors $\mathbf{X}_i$ are shown for clarity. The sketch shows a more general case where the the distance between neighboring spin centers is allowed to vary.}
	\label{fig:Figure1}
\end{figure}

The effect of the deformation on the energy of the ground total spin state $\left|S^2,M_s\right\rangle_g$ of the MM can be analyzed with the spin Hamiltonian $\hat{H}$. Before the deformation, $\hat{H} = -\sum_{(i,j)\in E} J \hat{\mathbf{s}}_i \cdot \hat{\mathbf{s}}_j $, where $E$ is the collection of edges in the molecule. Under the ferromagnetic coupling assumption the energy of $\left|S^2,M_s\right\rangle_g$ is $-n_e J$, where $n_e$ is the number of edges in the molecule and $J = J(r)$. The deformation adds a perturbation $\hat{H}' = -\sum_{(i,j)\in E} \Delta J_{ij} \hat{\mathbf{s}}_i \cdot \hat{\mathbf{s}}_j $ to $\hat{H}$.  The perturbation to the energy of $\left|S^2,M_s\right\rangle_g$ is therefore $-\sum_{(i,j) \in E}\Delta J_{ij}$. Now, consider the random variable $R=\sum_{(i,j)\in E}\Delta J_{ij}/(n_e J)$. To first-order, $R$ can be written as

\begin{equation}
	\centering R = \frac{c}{n_e J}\sum_{(i,j)\in E} \Delta r_{ij},
\end{equation}

\noindent
where $c = dJ(r+\Delta r)/d\Delta r|_{\Delta r = 0}$. $c$ measures the variation in the spin center-spin center coupling strength about the equilibrium separation. $R$ measures the size of the perturbation to the energy of $\left|S^2,M_s\right\rangle_g$ relative to the unperturbed energy of $\left|S^2,M_s\right\rangle_g$ under the random deformation. If $R$ is close to zero then the change in the energy of $\left|S^2,M_s\right\rangle_g$ is also small under the deformation. This means that if the	molecule	is initially	in state $\left|S^2,M_s\right\rangle_g$ then it will not make a transition out of $\left|S^2,M_s\right\rangle_g$ to another total spin state under the deformation. In other words, if $R$ is sufficiently small for a deformation, then the direction and magnitude of the total spin of a ground-state MM will not change under the deformation.\\

We say that the total spin of a particular MM is a \emph{weak topological invariant} if 

\begin{equation}
 P\left(|R| < \epsilon\right) > 1 - \epsilon
\end{equation}

\noindent
for some small $\epsilon > 0$, where $P$ is the probability measure (from the probability space on which $R$ is defined). In order to determine whether a particular structure for a MM has a robust total spin, we therefore need to compute the distribution of $R$ and verify whether equation (3) is satisfied. In the appendix it is shown that $R$ is a normal random variable with mean zero and variance

\begin{equation}
	\centering \mbox{var}(R)=2 \sigma^2_x (c/J)^2 Q,
\end{equation}

\noindent
where

\begin{equation}
\centering Q=\frac{1}{n_e} + \frac{1}{n_e^2} \sum_{\phi \in A}\cos\phi.
\end{equation}

\noindent
Here $A$ is the set of internal angles of the molecule (angles between adjacent edges). $Q$ is always positive because $\mbox{var}(R) \geq 0$ by definition. Because $R$ is a normal random variable, values of $P\left(|R| < \epsilon\right)$ that are close to 1 will be achieved for MMs that have small $\mbox{var}(R)$. Equation (4) therefore shows that the total spin of a MM will be stable to shape deformations if the factors $|c/J|$ and $Q$ are small. $|c/J|$ measures the degree to which the spin-spin interaction strength within the MM changes under an arbitrary deformation. There do not appear to be any experimental data or quantum chemical calculations that can be used to estimate a realistic value of $|c/J|$. We therefore set $|c/J| = 2$, which is reasonable because for small shape deformations the variation of $J$ about $r$ will be approximately quadratic. The factor $Q$ is related to the geometric structure of the MM. Equation (4) therefore allows us to explore how the geometry of a MM is related to the stability of the total spin through equation (5). For MMs containing more than several spins, this task is unsuitable for quantum chemical calculations with current computing power. \\

Let us consider the Platonic solids (the tetrahedron, octahedron, cube, dodecahedron and icosahedron) as an example calculation. The Platonic solids resemble many of the spin center arrangements that appear in real MMs. For example, the tetrahedral and cube spin arrangement appears in some Mn$_4$ and Ln$_4$ complexes \cite{{Liu2009, Liu2012, Nguyen2011}} and the octahedral spin arrangement appears in the Mn$_6$Br$_4$(Me$_2$dbm)$_6$ ion \cite{Aromi1999}. We can compute $P(|R| < \epsilon)$ using the formula

\begin{equation}
	P(|R| < \epsilon) = P(R < \epsilon) - P(R < -\epsilon).
\end{equation}

Using $\epsilon = 0.05$, $\sigma_x = 0.15$ (i.e., deformations that shift the spin centers a distance of 15 \% of the edge lengths on average) and the formula for $\mbox{var}(P)$ and $Q$ above, we find that $P(|R| < \epsilon) = 0.16$ for the tetrahedron ($n_e =6$), $P(|R| < \epsilon) = 0.31$ for the cube ($n_e$ =12), $P(|R| < \epsilon) = 0.23$ for the octahedron ($n_e =12$), $P(|R| < \epsilon) = 0.70$ for the dodecahedron ($n_e =30$), and $P(|R| < \epsilon) = 0.48$ for the icosahedron ($n_e = 30$). We would therefore expect that the total spin of a ferromagnetic MM with a tetrahedral arrangement of spins to be much more sensitive to shape deformations to the molecule than the total spin of a MM with a dodecahedral arrangement of spins. However, in none of these cases is $P(|R| < 0.05) > 0.95$, and therefore none of these molecules has a weak topological invariant total spin. \\

Before proceeding, let us establish the connection between the definition of a weak topological invariant and the mathematical concept of convergence in probability. Suppose we have an infinite sequence of MMs labelled by $1, 2, \ldots$, such that for all $\epsilon > 0$, $P\left(|R_k| < \epsilon\right) \rightarrow 1$. By definition, this means that the sequence of random variables $R_1, R_2, \ldots$ converges to zero in probability. Now, convergence in probability of the sequence $R_1, R_2, \ldots$ can be described by the Ky-Fan metric \cite{Durrett},

\begin{equation}
d(R_k,0) = \inf\left\{\delta > 0: P\left(|R_k| < \delta \right) > 1 - \delta \right\},
\end{equation}

\noindent
which shows that for any fixed $\epsilon > 0$, there exists an integer $n_0$ such that for all $k > n_0$, $P\left(|R_k| < \epsilon\right) > 1 - \epsilon$. Hence all MMs with labels greater than $n_0$ have weak topological invariant total spins. Moreover, if the rate of convergence in probability of the sequence $R_1, R_2, \ldots$ is known, then the integer $n_0$ can be evaluated and MMs with weak topological invariant total spins can be identified. We will make use of this connection further in section III. 

\section{2D Ferromagnetic MMs}

The connection between weak topological invariance and convergence in probability suggests that a better strategy for searching for MMs with weak topological invariant total spins is to set up a sequence of MMs such that $R \rightarrow 0$ in probability. We therefore consider the special case of 2D ferromagnetic MMs (2DFMM). Let $P_k$ denote a polygon with $k$ spin centers and constant edge lengths. $P_{k_1} | P_{k_2}$ indicates that the polygons $P_{k_1}$ and $P_{k_2}$ share an edge, and $(P_{k_1} | P_{k_2} )| P_{k_3}$ indicates that polygon $P_{k_3}$ shares an edge with any one of the edges of the structure $P_{k_1} | P_{k_2}$ . An arbitrary 2DFMM can be constructed in the following way. Choose $N$ polygons $P_{k_1} ,P_{k_2},...,P_{k_N}$ and create the structure $P_{k_1} | P_{k_2}$. Then create the structure $(P_{k_1} | P_{k_2} )| P_{k_3}$, and then the structure $((P_{k_1} | P_{k_2} )| P_{k_3})| P_{k_4}$, and so on. The resulting structure has the form (Figure 3)

\begin{equation}
	\centering T = (((P_{k_1}|P_{k_2})|...)|P_{k_{N-1}})|P_{k_N}.
\end{equation}	

\begin{figure}
	\centering
		\includegraphics[width=0.50\textwidth]{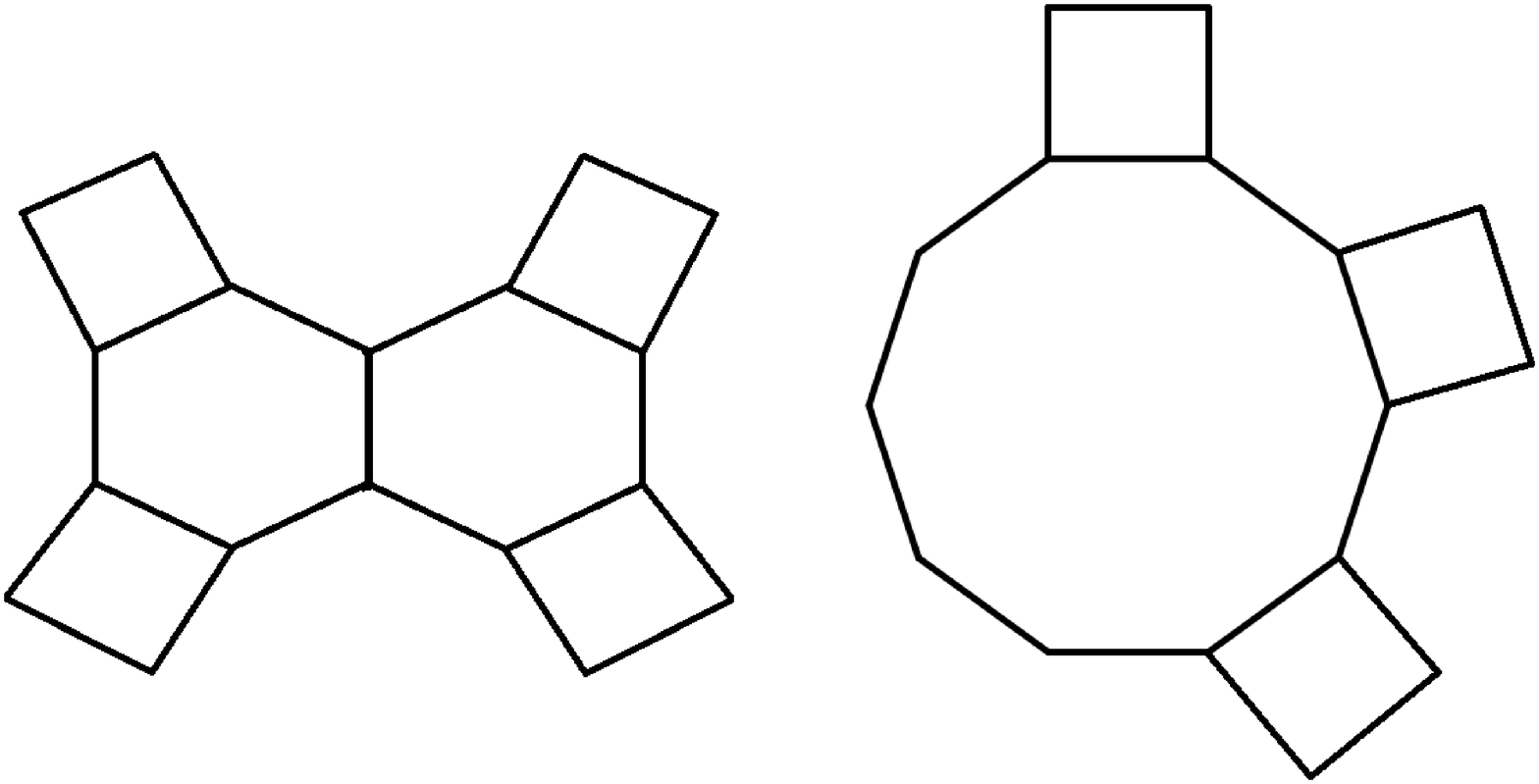}
	\caption{Examples of 2D ferromagnetic MMs. The spin centers lie at the vertices of the polygons. The MM on the right is the maximal structure for the class $(n_v, N) = (16,4).$}
	\label{fig:Figure3}
\end{figure}

\noindent
The polygons $P_{k_1} ,P_{k_2},...,P_{k_N}$ of $T$ are required to all lie within the same plane, and the internal angles of each polygon are fixed at $\pi-2\pi/k$, where $k$ is the number of spins in the polygon. Note that the atoms of the ligands connecting the spin centers are not restricted to lie within the two-dimensional plane of the spin centers. We further require that a vertex in $T$ can only belong to two polygons at most. This means that for any three polygons $P_i$, $P_j$, and $P_l$ in a 2DFMM structure such that $P_i|P_j|P_l$, $P_i$ and $P_l$ are separated by at least one edge. This assumption is reasonable, because for 2DFMMs that do not satisfy this condition the ligands may be very close together or overlapping in space. Several examples of real MMs that have nearly planar geometries have been reported in the literature, including an Fe$_4$ complex \cite{Gatteschi1999NEW}, an Fe$_8$ complex \cite{Wiegardt1997NEW}, and an  Fe$_{19}$ complex \cite{Powel1999NEW}.

\subsection{Convergence in probability of sequences of 2DFMMs}

We can create a sequence of 2DFMMs as follows. First, note that an arbitrary 2DFMM can be classified according $(n_v, n_e, N)$, where $n_v$ is the number of vertices, $n_e$ is the number of edges, and $N$ is the number of polygons. Applying Euler's formula to a 2DFMM gives the result $n_e = n_v + N - 1$, which shows that the value of $n_e$ is known if both $n_v$ and $N$ are specified. We can therefore use the two numbers $(n_v, N)$ to classify the 2DFMMs. Now, fix a value of $N$ (say, $N_0$) and for each class $(n_v, N)$ with $N = N_0$, choose the 2DFMM that has the largest value of $Q$ in equation (5). For any fixed $N$, we can therefore construct an infinite sequence of 2DFMMs such that the number of vertices $n_v$ increases along the sequence. Our next goal is to show that for such a sequence, $R \rightarrow 0$ in probability as $n_v \rightarrow \infty$. In the following, a subscript $(n_v,N)$ on a quantity means that the quantity pertains a 2DFMM from the class ${(n_v,N)}$.

For a given class $(n_v,N)$, let us first identify a 2DFMM that has the largest value of $Q_{(n_v,N)}$. Because $n_e$ is fixed for a 2DFMM in the class $(n_v,N)$, it is sufficient to consider the quantity 

\begin{equation}
\Phi_{(n_v,N)} = \sum_{\phi \in A}\cos\phi
\end{equation}

\noindent
from equation (5). We propose that a structure of the type shown on the right hand side of Figure 3 (the `maximal structure') has the largest value of $\Phi_{(n_v,N)}$ of all 2DFMMs in the class $(n_v,N)$. Let the notation $P_{k_i}|cP_{k_j}$ indicate that the polygon $P_{k_i}$ shares edges with $c$ polygons of the form $P_{k_j}$, and let $P_k^c=P_k|P_k|...P_k$, where the polygon $P_k$ appears $c$ times in the structure $P_k|P_k|...P_k$.  The maximal structure for the class $(n_v,N)$ can be written as

\begin{equation}
	\centering T_{(n_v,N)}^*=\begin{cases}
P_{n_v-2(N-1)}|(N-1)P_4 &  \mbox{ if } (N-1)-A_{(n_v,N)}\leq 0\\
(A_{(n_v,N)}-1)P_4|P_{n_v-2(N-1)}|P_4^{N-A_{(n_v,N)}} &  \mbox{ if } (N-1)-A_{(n_v,N)}> 0,
	\end{cases}
\end{equation}
\vspace{3 mm}

\noindent
where $A_{(n_v,N)}=(n_v-2(N-1))/2$ if $n_v -2(N-1)$ is even and $A_{(n_v,N)}=(n_v-2(N-1) - 1)/2$ if $n_v -2(N-1)$ is odd. $A_{(n_v,N)}$ is the maximum number of polygons that can share an edge with the polygon $P_{n_v-2(N-1)}$. The factor $\Phi_{(n_v,N)}$ for the maximal structure works out to be

\begin{equation}
	\centering\Phi_{(n_v,N)}^*=(n_v-2(N-1))\cos\frac{2\pi}{n_v-2(N-1)}-2(N-1)\sin\frac{2\pi}{n_v-2(N-1)},
\end{equation}
\vspace{3 mm}

\noindent
if $(N-1) - A_{n_v,N} \leq 0$, and

\begin{equation}
	\centering\Phi_{(n_v,N)}^*=(n_v-2(N-1))\cos\frac{2\pi}{n_v-2(N-1)}-2A_{(n_v,N)}\sin\frac{2\pi}{n_v-2(N-1)}-2(N-A_{(n_v,N)}-1),
\end{equation}

\noindent
if $(N-1) - A_{n_v,N} > 0$. Our proposition that $\Phi_{(n_v,N)}^*\geq\Phi_{(n_v,N)}$ for any 2DFMM in the class $(n_v,N)$ was	confirmed	via Monte	Carlo simulations. In these simulations, we generated $10^5$ independent 2DFMMs with $N$ polygons, where $N$ was uniformly sampled between 2 and 100 and $k_1,\ldots,k_N$ were uniformly sampled between 4 and 10.  We also performed simulations of $10^3$ independent 2DFMMs with $k_2,k_4,...$ sampled between 4 and 5 and $k_3,k_5$ sampled between 6 and 10 to explore the effect of combinations of small and large neighboring polygons. In every simulation, it was found that $\Phi_{(n_v,N)}$ was always equal to or less than $\Phi_{(n_v,N)}^*$ for the maximal structure $T_{(n_v,N)}^*$.  For any 2DFMM from the class $(n_v,N)$, we can therefore estimate $Q_{(n_v,N)}$ with the inequality

\begin{equation}
	\centering Q_{(n_v,N)}\leq Q_{(n_v,N)}^* = \frac{1}{n_v+N-1}+\frac{\Phi_{(n_v,N)}^*}{(n_v+N-1)^2}.
\end{equation}
\vspace{3 mm}

\noindent
Moreover, we can estimate the probability $P\left(|R_{(n_v,N)}| < \epsilon\right)$ for any 2DFMM from the class $(n_v,N)$ with the inequality

\begin{equation}
P(|R_{(n_v,N)}| < \epsilon) \geq P(Z_{(n_v,N)} < \epsilon) - P(Z_{(n_v,N)} < -\epsilon),
\end{equation}

\noindent
where $Z_{(n_v,N)}$ is a Gaussian random variable with mean 0 and variance $\mbox{var}(R)=2\sigma_x^2(c/J)^2 Q_{(n_v,N)}^*$. The bound in equations (13) and (14) are the tightest bounds to $Q_{(n_v,N)}$ and $P\left(|R_{(n_v,N)}| < \epsilon\right)$ that can be constructed for an arbitrary 2DFMM from the class $(n_v,N)$. Proving equations (13) and (14) rigorously is very difficult, so we will not attempt this here.\\

Let us now fix a value of $N$ and consider the sequence of 2DFMMs with increasing $n_v$ that was constructed previously. The previous paragraph shows that this is actually a sequence of maximal structures. The right-hand side of equation (13) shows that $Q_{(n_v,N)}^* \rightarrow 0$ as $n_v \rightarrow \infty$ for any fixed $N$. Now, let us consider \emph{any} infinite sequence of 2DFMMs (not necessarily maximal structures) such that the number of vertices $n_v$ is increasing to infinity and $N$ is fixed. The upper-bound in equation (13) then shows that $Q_{(n_v,N)} \rightarrow 0$ as $n_v \rightarrow \infty$. Equation (4) and the Markov inequality then shows that for this sequence $P(|R_{(n_v,N)}|<\epsilon)$ also converges to 1 for all $\epsilon > 0$, and hence that $R_{(n_v,N)} \rightarrow 0$ in probability (note that this is not a trivial consequence of the law of large numbers, because the random variables in the numerator of equation (2) are not independent). The definition in equation (3) then shows that the total spin of any sufficiently large 2DFMM is a weak topological invariant.

\subsection{Predictions of 2DFMMs with weak topological invariant total spins}

\begin{figure}
	\centering
		\includegraphics[width=0.40\textwidth]{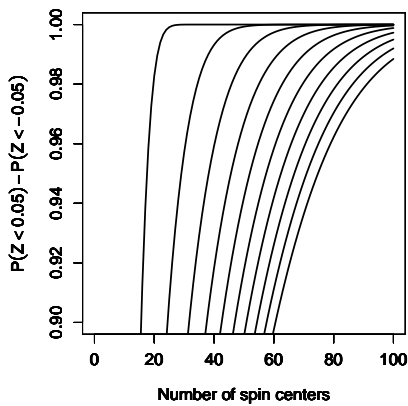}
	\caption{Plots of $P(Z_{(n_v,N)} < \epsilon) - P(Z_{(n_v,N)} < -\epsilon)$ for $\epsilon = 0.05$. Starting from the left, the curves correspond to $N = 1$ (one spin ring) up to $N = 10$ (ten spin rings) in order. The curves are tight lower-bounds to $P(|R_{(n_v,N)}| < \epsilon)$ for any arbitrary 2DFMM with $n_v$ spin centers and $N$ spin rings. The regions where $P(Z_{(n_v,N)} < \epsilon) - P(Z_{(n_v,N)} < -\epsilon) > 0.95$ are where the total spin is a weak topological invariant.}
	\label{fig:Figure3}
\end{figure}

The above construction shows that for any arbitrary 2D ferromagnetic MM with $n_v$ spin centers and $N$ polygons, $P(|R| < \epsilon)$ is bounded below according to equation (14). We can use this to estimate the number of spin centers that a 2DFMM must have for the total spin to be a weak topological invariant. A plot of $P(Z_{(n_v,N)} < \epsilon) - P(Z_{(n_v,N)} <- \epsilon)$ is shown in Figure 4 using $\sigma_x = 0.15$ and $\epsilon=0.05$ for various $n_v$ and $N$. When $P(Z_{(n_v,N)} < \epsilon) - P(Z_{(n_v,N)} <- \epsilon)$ exceeds the value 0.95 then the total spin of any 2DFMM with $n_v$ spin centers and $N$ polygons can be regarded as a weak topological invariant. It can be seen that for $N$ between 1 and 5 only between 20 and 50 spins are needed for the total spin of a 2DFMM to be a weak topological invariant, whereas for $N$ between 6 and 10 around 60 to 80 spins are needed for the total spin to be a weak topological invariant. This calculation therefore shows that the stability of the total spin of a 2DFMM is enhanced when the number of rings in the molecule is small. Some examples of 2DFMMs with weak topological invariant total spins are shown in Figure 5. 2DFMMs such as these containing around 20 - 50 spins are much smaller than the largest MMs that have been reported in the literature (such as the Mn$_{84}$ torus \cite{AromiBrechin2006,Tasiopoulos2004}), and are therefore 2DFMMs of this size built from a small number of spin rings (between 1 and 5) may be reasonable targets for synthetic chemistry. A particularly stable total spin can be achieved when the molecule contains only one spin ring. Smaller molecules of this type consisting of eight to ten metal ions have been widely studied in the literature \cite{Fe10, Cr8}. Slightly larger spin rings consisting of 20 ions therefore appear to be very reasonable targets for the synthesis of MMs with weakly topological invariant total spins.  \\

\begin{figure}
	\centering
		\includegraphics[width=0.50\textwidth]{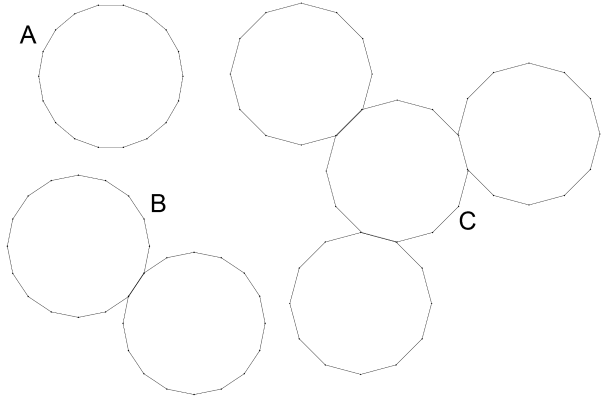}
	\caption{Examples of 2DSFMMs with weak topological invariant total spins for $\epsilon = 0.05$. A. $n_v = 18, N = 1$. B. $n_v = 30, N = 2$. C. $n_v = 42, N = 4$ (C). }
	\label{fig:Figure4}
\end{figure}

\section{Discussion and Final Remarks}

By using a modification of the spin Hamiltonian model, we studied how the energy of the ground total spin state of a MM varies under small, random shape deformations to the molecule. While we have focused on the idealised case of ferromagnetic MMs, our model can also be applied to other cases by allowing for the coupling constants and distances between pairs of spins to vary. However, the advantage of studying the ferromagnetic case is that it straightforwardly leads to molecular magnets with highly stable magnetic moments. By employing mathematical concepts such as convergence in probability and convergence bounds, we showed that for 2D ferromagnetic MMs containing a sufficiently large number of spins (between 20 and 50) and a small number of spin rings (between 1 and 5), the energy of the ground total spin state is extremely insensitive to shape deformations. These molecules are therefore very unlikely to make transitions out of the ground spin state under a deformation, and the magnitude and direction of the total spin will remain unchanged. In this sense we can regard the total spin of such MMs as a `weak topological invariant', i.e., a quantity that only depends upon the arrangement of the spin centers in the molecule and not upon shape deformations to the molecule. The prefix `weak' means that there may exist deformations that cause a large change in the total spin of the molecule, but which have a low probability of occurring. The result holds for `small' shape deformations. While there is no clear upper-bound on the size of the deformations, computer simulations of our model for a variety of MMs show that equations (2) and (3) holds up to about $\sigma_x = 0.2$, i.e., deformations that shift the spin centers a distance of about 20 \% of the spin center-spin center distance (result not shown). For the case of the Mn$_{12}$(Ac) molecule the average distance between neighboring Mn ions is about 3.20 \AA\ . For 2DFMMs with edge lengths around 3.20 \AA\ , our conclusions should therefore hold for deformations that shift the spin centers up to a distance of about 0.60 \AA\ on average. The ionic radii of Mn$^{3+}$ and Mn$^{4+}$ are around 0.58 \AA\ and 0.53 \AA\ ,  respectively \cite{HandbookofCP}, which are comparable to the distance 0.60 \AA\ . Thus, by `small' deformations we mean deformations that shift the spin centers about one ionic radius away from their equilibrium position. 

While two-dimensional MMs containing 20 - 50 spin centers do not appear to have been reported in the literature, these sizes are well within the size range of the three-dimensional MMs that have been reported, and therefore might be good targets for chemical synthesis. The requirement that all spin-spin couplings be ferromagnetic poses an interesting challenge for synthetic chemists, because in a majority of the molecules synthesised to date the magnetic properties arise from a competition between both ferromagnetic and antiferromagnetic couplings. Moreover, it is known in MMs with anisotropy that the strength of the anisotropy scales roughly inversely with the number of spins \cite{PCCPNEW}. It might therefore turn out that large 2DFMMs with weak topological invariant magnetic moments lack some of the interesting physics that is seen in smaller MMs such as Mn$_{12}$. Nonetheless, these 2DFMMs may be attractive targets for applications which strictly require molecular-sized components with highly robust magnetic moments.

While our model and calculations appear reasonable, there does not appear to be any reasonable experimental data or quantum chemical calculation that we can compare our results to. While it is standard to perform density functional theory calculations on small MMs, these calculations are very time consuming and are not practical for the present problem, which requires the calculations to be repeated many times for many different deformations to the molecule. Moreover, these calculations are limited to MMs containing less than a dozen metal ions, whereas our claims hold for larger MMs. However, as long as one is content working with the spin Hamiltonian (which is widely regarded qualitatively accurate for MMs \cite{FriedmanSarachik, SessoliNature}) and small displacements, then the conclusions from our work are scientifically meaningful. An important issue that needs to be addressed is how the ratio $|c/J|$ in equations (2) and (4) varies as the spin center-spin center distance varies about its equilibrium distance. We interpreted our calculations by assuming that $J$ varies quadratically about $r$, which is expected if the variation is sufficiently small. However, it may be that $J$ varies in a much more complicated manner within the deformation range in which the model applies. For example, it is known that $J$ also depends on the Mn-O-Mn angles in various MMs \cite{Jangles} and the entire structure of the ligand connecting the two spins \cite{Bou10}. In a future study we will extract the functional form of $J$ from detailed first-principles calculations, and also consider applying similar mathematical approaches to the design of molecules with other highly functional properties.\\

\section{Appendix}

The distribution of $R$ in equation (2) can be computed as follows. Let $r_{ij}'$ be the distance between spin centers $i$ and $j$ after the spin centers are shifted by the random vectors. We have $r_{ij}' = \left|\left(\mathbf{r}_i + \mathbf{X}_i\right) - \left(\mathbf{r}_j + \mathbf{X}_j\right)\right|$. Expanding $r_{ij}'$ into its Taylor's series and retaining only the terms that are linear in the components of  $\mathbf{X}_i$ and $\mathbf{X}_j$ gives

\begin{eqnarray}
		\Delta r_{ij} &=& \frac{1}{r_{ij}} \left(\left(r_i^x - r_j^x\right)X_i + \left(r_i^y - r_j^y\right)Y_i \right. \nonumber \\
									&+& \left.\left(r_i^z - r_j^z\right)Z_i + \left(r_j^x - r_i^x\right)X_j + \left(r_j^y - r_i^y\right)Y_j + \left(r_j^z - r_i^z\right												)Z_j\right) 
	\end{eqnarray}

\noindent
where $\Delta r_{ij} = r_{ij}' - r_{ij}$, $\left(r_i^x,r_i^y,r_i^z\right) = \mathbf{r}_i$, $\left(r_j^x,r_j^y,r_j^z\right) = \mathbf{r}_j$, $\left(X_i,Y_i,Z_i\right) = \mathbf{X}_i$ and $\left(X_j,Y_j,Z_j\right) = \mathbf{X}_j$. $X_i,Y_i,Z_i,X_j,Y_j,Z_j$ are independent normal random variables with mean zero and variance $\sigma_x^2$. $r_{ij} = 1$ by assumption. Substituting equation (15) into equation (2) shows that $R$ is a normal random variable with mean zero. The variance can be computed with the formula

\begin{equation}
\mbox{var}(R) = \frac{(c/J)^2}{n_e^2}\left(\sum_{(i,j) \in E} \mbox{var}(\Delta r_{ij}) + 2\sum_{(i,j),(k,l) \ \in E^2, (i,j) \neq (k,l)} \mbox{cov}\left(\Delta r_{ij} \Delta r_{kl}\right)\right),
\end{equation}  

\noindent
where $E$ is the edge set of the molecule under consideration and $n_e$ is the number of edges in the molecule. Choose an arbitrary edge from the molecule $(i,j)$. Using equation (15), we find that 

\begin{equation}
\mbox{var}\left(\Delta r_{ij}\right) = 2\sigma_x^2\left(\left(r_i^x - r_j^x \right) + \left(r_i^y - r_j^y \right) + \left(r_i^z - r_j^z \right)\right). 
\end{equation}

\noindent
The term in the brackets is simply $r_{ij}^2 = 1$, and so the first sum in equation (16) is

\begin{equation}
\sum_{(i,j) \in E} \mbox{var}\left(\Delta r_{ij}\right) = 2n_e\sigma_x^2.
\end{equation}

\noindent
Choose an arbitrary pair of edges $(i,j)$ and $(k,l)$ from $E^2$. Because $\Delta r_{ij}$ and $\Delta r_{kl}$ both have mean zero, the covariance of $\Delta r_{ij}$ and $\Delta r_{kl}$ is equal to $E\left(\Delta r_{ij}\Delta r_{kl}\right)$. According to equation (15), if one of $k$ or $l$ not is equal to either $i$ or $j$, then $\Delta r_{ij}$ and $\Delta r_{kl}$ are independent random variables and $E\left(\Delta r_{ij}\Delta r_{kl}\right) = 0$. We therefore only need to consider the case where the edges $(i,j)$ and $(k,l)$ are adjacent in the molecule. Suppose that $j = l$. Using equation (15) and the fact that $E\left((X_j - X_i)(X_k - X_j)\right) = -\sigma_x^2$ (and similarly for the $y$ and $z$ components) we find that 

\begin{equation}
E\left(\Delta r_{ij} \Delta r_{jk}\right) = -\sigma_x^2\left(\left(r_j^x - r_i^x\right)\left(r_k^x - r_j^x\right) + \left(r_j^y - r_i^y\right)\left(r_k^y - r_j^y\right) + \left(r_j^z - r_i^z\right)\left(r_k^z - r_j^z\right)\right).
\end{equation}

\noindent
The term in the brackets is equal to $\left(\mathbf{r}_i - \mathbf{r}_j\right)\cdot\left(\mathbf{r}_j - \mathbf{r}_k\right) = \cos\left(\pi - \phi\right)$, where $\phi$ is the angle $i,j,k$. The second sum in equation (16) therefore works out to be

\begin{equation}
\sum_{(i,j),(k,l) \ \in E^2, (i,j) \neq (k,l)} \mbox{cov}\left(\Delta r_{ij} \Delta r_{kl}\right) = \sigma_x^2 \sum_{\phi \in A} \cos\phi,
\end{equation}  

\noindent
where $A$ is the set of internal angles in molecule (angles between adjacent edges). Substituting equations (20) and (18) into (16) gives equation (4) of the main paper.

\section*{Acknowledgements}
\noindent
This work was supported by the World Premier International Research Center Initiative (WPI), MEXT, Japan. H. G. K. ~acknowledges the National Science Foundation (Grant No. ~DMR-1151387). We thank Ross McDonald for helpful conversations. 

\bibliographystyle{PNAS2009}
\bibliography{packwood_5-6-13_PRSA_2}

\end{document}